\documentclass{article}
\usepackage[]{amsmath,amssymb}
\usepackage{graphics,epsfig}

\begin{document}

\date{}
\title{Analytic results on the geometric entropy for free fields}
\author{H. Casini\footnote{e-mail: casini@cab.cnea.gov.ar} 
 \, and M. Huerta\footnote{e-mail: huerta@cabtep2.cnea.gov.ar} \\
{\sl Centro At\'omico Bariloche,
8400-S.C. de Bariloche, R\'{\i}o Negro, Argentina}}
\maketitle

\begin{abstract}
The trace of integer powers of the local density matrix $\rho_V$ corresponding to the vacuum state reduced to a region $V$ can be formally expressed in terms of a functional integral on a manifold with conical singularities. Recently, some progress has been made in explicitly evaluating this type of integrals for free fields. However, finding the associated geometric entropy remained in general a difficult task involving an analytic continuation in the conical angle. In this paper, we obtain this analytic continuation explicitly exploiting a relation between the functional integral formulas and the Chung-Peschel expressions for $\rho_V$ in terms of correlators. The result is that the entropy is given in terms of a functional integral in flat Euclidean space with a cut on $V$ where a specific boundary condition  is imposed. As an example we get the exact entanglement entropies for massive scalar and Dirac free fields in $1+1$ dimensions in terms of the solutions of a non linear differential equation of the Painlev\'e V type.
\end{abstract}

\section{Introduction}
Recently, there has been a growing interest in different areas of physics for understanding the manifestations of quantum entanglement in a variety of phenomena. Several measures of entanglement have been studied either as a tool to analyze these phenomena or as a subject of investigation in itself. 
The applications range from condensed matter physics \cite{cm} and quantum field theory \cite{qft} to the physics of black holes \cite{bh} and holography in the context of string theory \cite{japoholo}. 

In quantum field theory the typical manifestation of entanglement is the statistical correlation of measurements  for  independent (commuting) sets of observables located at spatially separated regions.  A natural measure of the entanglement between  a region $V$ with the rest of the space is the geometric entropy \cite{geoent}, also known as entanglement entropy. This is defined as the von Neumann entropy $S(V)=-\textrm{tr}(\rho_V\log \rho_V)$ of the reduced density matrix of the vacuum state  $\rho_V=\textrm{tr}_{-V}\left|0\right\rangle\left\langle 0\right|$ in the set $V$ (the trace in the expression for $\rho_V$ is taken over the degrees of freedom lying outside $V$). 
 
More generally, there is a one parameter family of information measures also associated to $\rho_V$ known as R\'enyi entropies (or alpha-entropies),
$ S_n(V)=\log(\textrm{tr}\rho_V^n)/(1-n) ,\label{cuatro}$
satisfying the following particular limit 
\begin{equation}
S(V)=\lim_{n\rightarrow 1} S_n (V)\,.
\label{limit}
\end{equation}
These behave very much like the entanglement entropy. For integer $n$ they are also more suitable to explicit calculation, since the traces $\textrm{tr}\rho_V^n$ involved in (\ref{cuatro}), with $n\in
{\mathbb Z}$, can be represented by a functional integral on an $n$-sheeted $d$ dimensional Euclidean space with conical singularities located at the boundary of the set $V$ \cite{conicas}. Here the functional integral is defined on one dimension more than the one corresponding to $V$, which is $d-1$. The replicated space is obtained considering $n$ copies of the $d$-space cut along $V$, and sewing together the upper side of the cut in the $k^{\textrm{th}}$ copy with
the lower one of the $(k+1)^{\textrm{th}}$ copy, for $ k=1,...,n$, and where the copy $n+1$ coincides with the first one.
The trace of $\rho^{n}$ is then given by the functional integral
$Z_n$ for the field in this manifold, 
\begin{equation}
\textrm{tr}\rho^{n}= \frac{Z_n}{(Z_1)^{n}}\,. 
\label{dd}
\end{equation}
In general, solving this integral explicitly is a very difficult problem since we have to deal with a non trivial manifold resulting from the replication method. In this scheme, the entanglement entropy follows from $S_n$ by analytic continuation to $n=1$. 
In the case of free fields, a great simplification follows by mapping the $n-$sheeted problem to an equivalent one in which we deal with $n$ decoupled and multivalued free fields \cite{fermion}. First, we arrange the values of the field in the different copies in a single vector field $\vec{\Phi}$ living in a $d$ dimensional space,
\begin{equation}
\vec{\Phi}=\left(\begin{array}{c} 
\phi _{1}(\vec{x}) \\ 
\vdots \\  
\phi _{n}(\vec{x}) \end{array}
\right) \,,
\end{equation}
where $\phi _{l}(\vec{x})$ is the field on the $l^{\textrm{th}}$ copy. 
Note that in this way the space is simply connected but the singularities at the boundaries of $V$ are still there since the vector $\vec{\Phi}$ is not singled valued.  In fact, crossing $V$ from above or from below, the field gets multiplied by a matrix $T$ or $T^{-1}$ respectively. The matrix $T$ is given by
\begin{equation}
\begin{array}{c}
T=\left(
\begin{array}{lllll}
0 & 1 &  &  &  \\
& 0 & 1 &  &  \\
&  & . & . &  \\
&  &  & 0 & 1 \\
(\pm1)^{n+1} &  &  &  & 0
\end{array}
\right)
\end{array}\,, 
\end{equation}
where the upper sign corresponds to the bosonic case and the lower one to fermions \cite{fermion,boson,kabat}. This matrix has eigenvalues $e^{i\frac{k}{n}2\pi }$, with
$k_s=0\,,...,\,(n-1)$ or $k_f=-(n-1)/2\,,...,\,(n-1)/2$ depending on the bosonic or fermionic character of the field. 
Then, changing basis by a unitary transformation in the replica space,
we can diagonalize $T$, and the problem is reduced to $n$ decoupled
fields $\tilde{\phi}_{k}$ living on a single $d$ dimensional space. These fields are multivalued and live on the euclidean $d$ dimensional space with boundary conditions imposed on the $d-1$ dimensional set $V$ given by
\begin{equation}
\tilde{\phi}^{(+)}_k(\vec{r})=e^{i\frac{2 \pi k}{n}}\tilde{\phi}^{(-)}_k(\vec{r})\,,\,\,\,\,\,\,\,\,\,\,\, \, \vec{r}\in V\,.
\label{bc}
\end{equation}
Here $\tilde{\phi}^{(+)}_k$ and $\tilde{\phi}^{(-)}_k$ are the limits of the field as the variable approaches $V$ from each of its two opposite sides in $d$ dimensions. 
In this formulation we have for fermions~\footnote{In previous work \cite{fermion,boson,angulo} we have used the notation $Z[a]$ for the present $Z[e^{i 2 \pi a}]$.}  
\begin{equation}
\log(\textrm{tr}\rho_V^n)=\sum_{k=-(n-1)/2}^{(n-1)/2}\textrm{log}\, Z\left[e^{i 2 \pi \frac{k}{n}}\right]\,,
\label{r2}
\end{equation}
and for bosons
\begin{equation}
\log(\textrm{tr}\rho_V^n)=\sum_{k=0}^{n-1}\textrm{log}\, Z\left[e^{i 2 \pi \frac{k}{n}}\right]\,,
\label{r1}
\end{equation}
where we have written $Z[\lambda]$ the partition function corresponding to a field on a single copy of the Euclidean space, which is multiplied by a factor $\lambda$ when the variable crosses $V$. A related construction for two dimensional integrable theories where $\textrm{tr} \rho^n$ is expressed in terms of correlators of twisting operators is shown in \cite{cc}. 
 
Explicit calculation of $Z\left[e^{i 2 \pi k/n}\right]$ has been carried out for several cases of interest \cite{fermion,boson,angulo}.  However, the evaluation of the entropy is limited by the difficulties in getting the analytic continuation of the sum for non-integer $n$ in order to take the limit (\ref{limit}).

In this paper we solve this problem by exploiting the relation between (\ref{r2}) and (\ref{r1}) with the expressions obtained by Chung and Peschel \cite{chung} for $\rho _{V}$  in
terms of correlators for free boson and fermion discrete systems. In the continuum limit, these provides the necessary tools to compute the entropy. We find that it can also be written directly in terms of a functional integral in flat space. The result is 
\begin{equation}
S=\int_1^{\infty}d\lambda \frac{2}{(\lambda-1)^2}\log Z\left[\lambda\right]\,,\label{vein}
\end{equation}
for free fermions, and 
\begin{equation}
S=\int_1^{\infty}d\lambda \frac{2}{(\lambda+1)^2}\log Z\left[-\lambda\right]\,,\label{boti}
\end{equation}
for free bosons. Note that the functional integral involved has a boundary condition on the cut $V$ given by a real factor (instead of a phase factor as in (\ref{r2}) and (\ref{r1})), being a positive factor for fermions and a negative one for  bosons,
\begin{equation}
\phi^{(+)}(\vec{r})=\pm \lambda \,\,\phi^{(-)}(\vec{r})\,, \hspace{1.5cm} \vec{r}\in V\,,\,\,\,\,\, \lambda \ge 0.
\end{equation}
In this new scheme, the calculation of the geometric entropy in the set $V$ and in any dimensions is reduced to the one of the corresponding functional  $Z[\lambda]$. 

In the next Section we derive these equations and, as byproduct, we obtain an interesting relation between the $d$-dimensional free energy $\textrm{log}\, Z[\lambda]$ and the resolvent of a specific correlator acting in the $d-1$ dimensional region $V$. 
In Section III we also show how the formulas for the entropy can be obtained directly from the expressions (\ref{limit}), (\ref{r2}) and (\ref{r1}), modulo some assumptions in the behavior of the partition function $Z[\lambda]$ as a function of $\lambda$. An example is given in Section IV where we compute the entanglement entropy corresponding to a segment of one dimensional quantum bosonic and fermionic chains which in the continuum limit are described by two dimensional scalar and Dirac massive field theories. 
 
\section{Geometric entropy and two point functions for free fields}
\subsection{Fermionic fields}

In \cite{chung} an expression for $\rho _{V}$ was given in terms of correlators for free boson and fermion discrete systems. In order to describe the continuum, consider first a set of fermionic fields $\Psi_k(x)$ which satisfy the canonical anticommutation relations
\begin{equation}
\left\{\Psi^{\dagger}_i(x),\Psi_j(y)\right\}=\delta(x-y)\delta_{ij}\,.
\end{equation}
The two point correlation functions restricted to $V$ are given by
\begin{equation}
\textit{C}_V^{ij}(x,y)=\left\langle0 \right|\Psi_i^{\dagger}(x)\,\Psi_j(y)\left|0\right\rangle\,\,\,\,\,;\,\,\,\,x,y\in V\,.
\end{equation}
By definition, the vacuum expectation value $\left\langle O_{V}\right\rangle $ for any
operator $O_{V}$ localized inside a region $V\,$ must coincide with tr$(\rho
_{V}O_{V})$, where $\rho _{V}\,$is the local density matrix. 

On the other hand, due to the Wick theorem any
correlator inside $V$ can be written in terms of $C_V^{ij}(x,y)$. The expectation
values calculated with $\rho _{V}$ must also satisfy Wick's theorem. This fixes 
$\rho _{V}$ to be of the form 
\begin{equation}
\rho _{V}=e^{-H}\,,
\end{equation}
where $H$ is a local Hamiltonian (modular Hamiltonian) which is quadratic on the fields inside the region $V$,
\begin{equation}
\textit{H}=\int_V dxdy \,\Psi_i^{\dagger}(x)H_{ij}(x,y) \Psi_j(y)\,\,\,\,;\,\,\,\,\,H^{*}_{ji}(y,x)=H_{ij}(x,y)\,.
\end{equation}
The relation $\textrm{tr}(\rho_V \Psi_i^{\dagger}(x)\,\Psi_j(y))=C_V^{ij}(x,y)$ gives the kernel $H_{ij}(x,y)$ as (we do not write the subscript $V$ for notational convenience)
\begin{equation}
H^T_{ij}(x,y)=\left[\log(1-C)-\log(C)\right]_{ij}(x,y)\,.
\end{equation}
Then we have
\begin{equation}
S=-\textrm{tr}\left[(1-C)\log(1-C)+C\log(C)\right]\,,
\end{equation}
and also
\begin{equation}
\log \left(\textrm{tr}\rho^n\right)=\log\det\left[(1-C)^n+C^n\right]=\sum_{k=-\frac{n-1}{2}}^{\frac{n-1}{2}}
\log\det\left(C-\frac{e^{i 2 \pi k/n}}{e^{i 2 \pi k/n}-1}\right)\,,
\label{autoval}
\end{equation}
where this last sum is over the roots of the polynomial $(1-C)^n+C^n$.
Recalling eq. (\ref{r2}) and comparing it with (\ref{autoval}), we conclude that
\begin{equation}
\log Z[\lambda]=\textrm{tr}\log(C-\lambda/(\lambda-1))\,,\label{tyty}
\end{equation} 
for any $\lambda$.

In order to proceed, we use an integral representation of the logarithm written in terms of the resolvent
\begin{equation}
\log C=-\int_0^{\infty} d\beta \left[\frac{1}{C+\beta}-\frac{1}{\beta+1}\right]\,.
\end{equation}
This gives for the entropy $S$ up to a non universal constant term 
\begin{equation}
S=-\int_0^{\infty}d\beta \,\beta\, \textrm{tr}\left[\frac{1}{C+\beta}-\frac{1}{C-(\beta+1)}\right] \label{ss}\,.
\end{equation}
From equation (\ref{tyty}) we have 
\begin{equation}
\textrm{tr}\frac{1}{C-\lambda/(\lambda-1)}=(\lambda-1)^2 \frac{d\log Z[\lambda]}{d\lambda}     \,.
\label{e2}
\end{equation}
Finally, we use the equation (\ref{e2}) in (\ref{ss}), changing variables $\beta=-\lambda/(\lambda-1)$, $\lambda\in (0,1)$, in the first term of the integrand, and $\beta=1/(\lambda-1)$, $\lambda\in (1,\infty)$, in the second one. These two terms are equal once one takes into account that the reflection symmetry of the action on the axis containing $V$ gives place to 
\begin{equation}
\log Z[\lambda]=\log Z[1/\lambda]\,.\label{veintidos}
\end{equation}
 After integrating by parts we get our final expression for the entropy (\ref{vein}).
In fact, the corresponding boundary term is given by $\left.1/(\lambda-1)\log Z[\lambda]\right|_1^{\infty}$. It vanishes at $\lambda=1$ due to (\ref{veintidos}) (note also that there is no contribution to universal terms in $S$  from $\log Z[1]$ which is the partition function without a cut). At $\lambda\rightarrow \infty$ it also goes to zero since the free energy $\log Z[\lambda]$ increases at most powerlike in the angle variable $\sim\log (\lambda)$ (we have however no general proof of this behavior).

\subsection{Bosonic fields}

Following similar arguments we have for bosonic fields the following expressions \cite{chung}
\begin{equation}
S=\textrm{tr}\left[(C+\frac{1}{2})\log(C+\frac{1}{2})-(C-\frac{1}{2})\log(C-\frac{1}{2})\right]\,,\label{prima}\\
\end{equation}
with $C=\sqrt {X P}$ and $X$ and $P$ the vacuum field and momentum correlators inside $V$ given by
\begin{eqnarray}
X^{ij}(x,y) &=&\left\langle \phi _{i}(x)\phi _{j}(y)\right\rangle \,,\\
P^{ij}(x,y) &=&\left\langle \pi _{i}(x)\pi _{j}(y)\right\rangle\,.
\end{eqnarray}
We also have 
\begin{equation}
\log \left(\textrm{tr}\rho^n\right)=-\textrm{tr}\left[\log\left((C+\frac{1}{2})^n-(C-\frac{1}{2})^n\right)\right]=\sum_{k=0}^{n-1}\log\det\left(C-\frac{1}{2}\frac{e^{i 2 \pi \frac{k}{n}}+1}{e^{i 2 \pi \frac{k}{n}}-1}\right)\,, 
\label{autovals}
\end{equation}
where the last sum is over the roots of the polynomial $(C+\frac{1}{2})^n-(C-\frac{1}{2})^n$.
Comparing expressions (\ref{r1}) and (\ref{autovals}) we conclude that
\begin{equation}
\log Z[\lambda]=\textrm{tr}\log\left(C-\frac{1}{2}\frac{\lambda+1}{\lambda-1}\right)\,.\label{qq}
\end{equation} 
From (\ref{prima}) and the integral representation for the logarithm we have the entropy in terms of the resolvent 
\begin{equation}
S=-\int_0^{\infty}d\beta\, \beta \,\textrm{tr}\left[\frac{1}{C-\frac{1}{2}+\beta}-\frac{1}{C+\beta+\frac{1}{2}}\right]\,.\label{gto}
\end{equation}
Taking the derivative in  (\ref{qq}) we obtain for the resolvent
\begin{equation}
\textrm{tr}\frac{1}{C-(\lambda+1)/(2(\lambda-1))}=(\lambda-1)^2 \frac{d\log Z[\lambda]}{d\lambda}     \,.
\end{equation}
Finally, we change variables in (\ref{gto}) as $\beta=1/(1-\lambda)$, $\lambda\in (-\infty,1)$, in the first term  and $\beta=\lambda/(1-\lambda)$, $\lambda\in (0,1)$, in the second one, in order to get the entropy (\ref{boti}) after an integration by parts. We also have to use the reflection symmetry relation (\ref{veintidos}).  

\section{A different derivation}
We can get the above expressions for the entropy in a different way by the following trick. First we write the sum as a contour integral. Consider the following identity 
\begin{eqnarray}
\frac{1}{1-n}\sum_{k=-(n-1)/2}^{(n-1)/2} f[k/n]=\frac{1}{2\pi i (1-n)}\oint dz \left( \sum_{k=-(n-1)/2}^{(n-1)/2} \frac{1}{z-\frac{k}{n}}\right) f[z]=
\\ \frac{1}{2\pi i (1-n)} \oint dz \,\, n \left(\psi(1/2-n/2-nz) -\psi(n/2+1/2-nz) \right)  f[z] \,,
\end{eqnarray}
where $\psi$ is the digamma function and the integral must encircle the interval $[-1/2,1/2]$ on the real line. This formula provides an analytic extension of the sum for non-integer $n$ but we have to consider an integration contour which includes all the poles of the combination of digamma functions in the integrand for non integer $n$. These are located on the real line and are greater than $-1/2$. Thus the integration contour encircles the $[-1/2,\infty ]$ segment and we assume that $f[z]$ does not have singularities on it. Besides it must decrease fast enough at infinity. Otherwise we can multiply it by a cutoff factor $e^{- \lambda z}$ and make $\lambda\rightarrow 0$ at the end.  We have in the limit $n\rightarrow 1^+$ 

\begin{equation}
 \frac{1}{1-n}\sum_{k=-(n-1)/2}^{(n-1)/2} f[k/n]\rightarrow \frac{1}{2\pi i }\oint dz \,\, \left( -\frac{1}{z(n-1)}+\frac{1}{2z^2}+\frac{\pi^2}{\sin^2(\pi z)}-\psi^\prime(z) \right)  f[z] \,.\label{rete}
\end{equation}
Imposing the symmetry $f[z]=f[-z]$ and $f[z]\sim z^2$ for $z$ around the origin, the pole of the function within the brackets for $z=0$ do not contribute. Also the function $\psi^\prime(z)$ has its poles on the negative real axis for $z\le -1$. Thus we find that only the term proportional to $\sin^{-2}(\pi z)$ is relevant for the contour integral. Then, if there are no singularities of $f[z]$ for $\textrm{Re}[z]\ge 0$ we can choose the integration contour to be the imaginary axis. Using (\ref{rete}) with $f(z)\equiv \log(Z[e^{i 2 \pi z}])$ we  recover the expression (\ref{vein}) for the fermion field entropy. The bosonic case can be treated similarly.

\section{Example: Free
massive scalar and Dirac fields in two dimensions}

In two dimensions, the partition function $Z[\lambda]$  where $V$ is a single interval of length $r$ is determined by a non linear differential equation. Following \cite{fermion} and \cite{boson}, we introduce functions $w_D$ and $w_S$ for  a Dirac fermion and a real scalar, which are related to the respective partition functions $Z$ by
\begin{eqnarray}
r\frac{d}{dr}\log Z[e^{2 \pi b}]=w_D(b,t)\,,\\
r\frac{d}{dr}\log Z[-e^{2 \pi b}]=w_S(b,t)\,,
\end{eqnarray}
where $t=m\,r$ and $m$ is the field mass.
For a fixed $b$ these functions are given in terms of the same differential equation for both, the Dirac and the real scalar fields,
\begin{eqnarray}
w&=&-\int_{t}^{\infty}dy\,\,y\,\,u^{2}(y)  \,,\\
u^{\prime \prime }+\frac{1}{t}u^{\prime} &=&\frac{
u}{1+u^{2}}\left( u^{\prime}\right)
^{2}+u\left( 1+u^{2}\right) -\frac{4 b^{2}}{t^{2}}\frac{u}{1+u^{2}}\,. 
\end{eqnarray}
 
The Dirac and scalar cases differ however on the boundary conditions, 
\begin{eqnarray}
u_D(t) &\rightarrow &\frac{2}{\pi} \sinh (b \pi) K_{i 2 b} (t)\,\,\,\,\,\,\,\, \textrm{as} \,\,\,\, t\rightarrow \infty \,\\
u_{D}(t)&\rightarrow& -2 b \,(\log t+\,q_{D})\,\,\,\,\,\,\,\textrm{as}\,\,\,\,t\rightarrow 0 \\
u_{S}(t)&\rightarrow &\frac{2}{\pi} \cosh (b \pi) K_{i 2 b} (t)\,\,\,\,\,\,\,\, \textrm{as} \,\,\,\, t\rightarrow \infty \\
u_{S}(t) &\rightarrow & \frac{-1}{t \,(\log t+q_S)}\;\;\;\;\;\textrm{as}\;\;t\to 0\,,
\end{eqnarray}
where $q_D=-\log(2)+2\gamma_E+\frac{\psi[ib]+\psi[-ib]}{2}$ and $q_S=-\log(2)+2\gamma_E+\frac{\psi[1/2+ib]+\psi[1/2-ib]}{2}$ with $\gamma_E$ the Euler constant, $\psi[a]$ the digamma function and $K$ the modified Bessel function. The boundary conditions at the origin can be deduced from the connection formulas for this Painlev\'e equation \cite{conection} or directly from the Green function on the cut plane in the massless limit \cite{boson,myers}. 

The entropic $c$-function 
\begin{equation}
c(t)=r\frac{d}{dr}S(r)
\end{equation}
gives the universal part of the entropy $S(r)$ for a single interval in two dimensions \cite{cteor}. Analogously to the Zamolodchikov's C function, it is dimensionless and decreasing with $r$, and in the conformal limit takes the value $C_V/3$, with $C_V$ the Virasoro central charge.  
We have then 
\begin{eqnarray}
c_D(t)&=&\int_0^{\infty}db \frac{\pi}{\sinh(\pi b)^2}w_{D}(b,t)\,,\label{uuy}\\
c_S(t)&=&\int_0^{\infty}db \frac{\pi}{\cosh(\pi b)^2}w_{S}(b,t)\label{vvy}\,.
\end{eqnarray}
These analytic results can be easily evaluated with high precision  by solving the differential equation numerically. We have checked they are in perfect accord with previous lattice simulations for $c$ \cite{boson}. The leading short and long distance terms read
\begin{eqnarray}
c_D(t)\sim \frac{1}{3} -\frac{1}{3} t^2 \log^2 (t)\,\,\,\,\, \textrm{for}\,\, t\ll 1\,\,; \hspace{.4cm} c_D(t)\sim \frac{1}{2} t K_1(2 t)\,\,\,\,\, \textrm{for}\,\,t\gg 1\,, \\
c_S(t)\sim \frac{1}{3}+\frac{1}{2 \log (t)}\,\,\,\,\, \textrm{for}\,\,t\ll1\,\,; \hspace{.4cm} c_S(t) \sim \frac{1}{4}  t K_1(2 t)\,\,\,\,\, \textrm{for}\,\,t\gg 1\,.
\end{eqnarray} 
More subleading terms in the short distance expansion for $c_D(t)$ can be obtained straightforwardly from (\ref{uuy}) and the corresponding series solution of the differential equation \cite{fermion}. This converges up to $t=1$.

Large distance series expansions in terms of multiple integrals can be obtained from (\ref{uuy}-\ref{vvy}) using the form factor expansions of the partition function. Remarkably, it has been recently shown that the large distance leading term behavior $c(r)\simeq mr K_1(2mr)$ also holds for a large class of integrable interacting theories \cite{cc}.    

Other universal terms in different dimensions such as the vertex contribution to the entropy for polygonal sets in $2+1$ \cite{angulo}, or the leading universal contribution for a long and thin rectangle \cite{boson}, can be similarly obtained.

\end{document}